\documentclass{ws-p9-75x6-50}
\newcommand{\optbar}[1]{\shortstack{{\tiny (\rule[.4ex]{1em}{.1mm})}
  \\ [-.7ex] $#1$}}
\begin{document}
\newcommand{\nn}{\noindent}
\newcommand{\cs}{\mbox{$\clubsuit$}}
\newcommand{\nl}{\nonumber \\}
\newcommand{\hf}{\hfill}
\newcommand{\naive}{na$\ddot{\imath}$ve}
\newcommand {\oa} {\mbox{${\cal O}( \alpha)$}}
\newcommand {\ho} {\mbox{${\cal O}( \alpha^{2})$}}
\hyphenation{brems-strah-lung}
\def\ss{\footnotesize}
\def\SS{\footnotesize}
\def\sss{\scriptscriptstyle}
\def\barp{{\raise.35ex\hbox{${\sss (}$}}---{\raise.35ex\hbox{${\sss )}$}}}
\def\bdbarp{\hbox{$B_d$\kern-1.4em\raise1.4ex\hbox{\barp}}}
\def\bsbarp{\hbox{$B_s$\kern-1.4em\raise1.4ex\hbox{\barp}}}
\def\dbarp{\hbox{$D$\kern-1.1em\raise1.4ex\hbox{\barp}}}
\def\dbarp{\hbox{$D$\kern-1.1em\raise1.4ex\hbox{\barp}}}
\def\dcp{D^0_{\sss CP}}
\def\dbar{{\overline{D^0}}}
\def\ks{K_{\sss S}}
\newcommand{\xd}{x_d}
\newcommand{\xs}{x_s}
\newcommand{\bd}{B_d^0}
\newcommand{\bdb}{\overline{B_d^0}}
\newcommand{\bs}{B_s^0}
\newcommand{\bsb}{\overline{B_s^0}}
\newcommand{\bu}{B_u^\pm}
\newcommand{\bsbar}{\overline{B_s^0}}
\newcommand{\beq}{\begin{equation}}
\newcommand{\eeq}{\end{equation}}
\newcommand{\absvcb}{\vert V_{cb}\vert}
\newcommand{\absvub}{\vert V_{ub}\vert}
\newcommand{\absvtd}{\vert V_{td}\vert}
\newcommand{\absvts}{\vert V_{ts}\vert}
\newcommand{\abseps}{\vert\epsilon\vert}
\newcommand{\epsp}{\epsilon^\prime/\epsilon}
\newcommand{\fbb}{f^2_{B_d}\hat{B}_{B_d}}
\newcommand{\fbbs}{f^2_{B_s}\hat{B}_{B_s}}
\newcommand{\fbd}{f_{B_d}}
\newcommand{\fbs}{f_{B_s}}
\newcommand{\fds}{f_{D_s}}
\def\rly#1{\mathrel{\raise.3ex\hbox{$#1$\kern-.75em\lower1ex\hbox{$\sim$}}}}
\def\lsim{\rly<}

\def \zpc#1#2#3{{\it Z.~Phys.,} C#1 (19#2) #3}
\def \plb#1#2#3{{\it Phys.~Lett.,} B#1 (19#2) #3}
\def \ibj#1#2#3{~#1, (19#2) #3}
\def \prl#1#2#3{{\it Phys.~Rev.~Lett.,} #1 (19#2) #3}
\def \prd#1#2#3{{\it Phys.~Rev.,} D#1 (19#2) #3} 
\def \npb#1#2#3{{\it Nucl.~Phys.}, B#1 (19#2) #3}
\def\ijmp#1#2#3{{\it Int.\ J.\ Mod.\ Phys.} {\bf A#1} (19#2) #3}
\def \stone{{\it B Decays}, edited by S. Stone (World Scientific, Singapore,
1994)}
\newread\epsffilein 
\newif\ifepsffileok 
\newif\ifepsfbbfound 
\newif\ifepsfverbose 
\newdimen\epsfxsize 
\newdimen\epsfysize 
\newdimen\epsftsize 
\newdimen\epsfrsize 
\newdimen\epsftmp 
\newdimen\pspoints 
\pspoints=1bp 
\epsfxsize=0pt 
\epsfysize=0pt 
\def\epsfbox#1{\global\def\epsfllx{72}\global\def\epsflly{72}%
 \global\def\epsfurx{540}\global\def\epsfury{720}%
 \def\lbracket{[}\def\testit{#1}\ifx\testit\lbracket
 \let\next=\epsfgetlitbb\else\let\next=\epsfnormal\fi\next{#1}}%
\def\epsfgetlitbb#1#2 #3 #4 #5]#6{\epsfgrab #2 #3 #4 #5 .\\%
 \epsfsetgraph{#6}}%
\def\epsfnormal#1{\epsfgetbb{#1}\epsfsetgraph{#1}}%
\def\epsfgetbb#1{%
%
%
\openin\epsffilein=#1
\ifeof\epsffilein\errmessage{I couldn't open #1, will ignore it}\else
%
%
 {\epsffileoktrue \chardef\other=12
 \def\do##1{\catcode`##1=\other}\dospecials \catcode`\ =10
 \loop
 \read\epsffilein to \epsffileline
 \ifeof\epsffilein\epsffileokfalse\else
%
%
 \expandafter\epsfaux\epsffileline:. \\%
 \fi
 \ifepsffileok\repeat
 \ifepsfbbfound\else
 \ifepsfverbose\message{No bounding box comment in #1; using defaults}\fi\fi
 }\closein\epsffilein\fi}%
%
%
\def\epsfclipstring{}
\def\epsfclipon{\def\epsfclipstring{ clip}}%
\def\epsfclipoff{\def\epsfclipstring{}}%
\def\epsfsetgraph#1{%
 \epsfrsize=\epsfury\pspoints
 \advance\epsfrsize by-\epsflly\pspoints
 \epsftsize=\epsfurx\pspoints
 \advance\epsftsize by-\epsfllx\pspoints
%
%
 \epsfxsize\epsfsize\epsftsize\epsfrsize
 \ifnum\epsfxsize=0 \ifnum\epsfysize=0
 \epsfxsize=\epsftsize \epsfysize=\epsfrsize
 \epsfrsize=0pt
%
%
 \else\epsftmp=\epsftsize \divide\epsftmp\epsfrsize
 \epsfxsize=\epsfysize \multiply\epsfxsize\epsftmp
 \multiply\epsftmp\epsfrsize \advance\epsftsize-\epsftmp
 \epsftmp=\epsfysize
 \loop \advance\epsftsize\epsftsize \divide\epsftmp 2
 \ifnum\epsftmp>0
 \ifnum\epsftsize<\epsfrsize\else
 \advance\epsftsize-\epsfrsize \advance\epsfxsize\epsftmp \fi
 \repeat
 \epsfrsize=0pt
 \fi
 \else \ifnum\epsfysize=0
 \epsftmp=\epsfrsize \divide\epsftmp\epsftsize
 \epsfysize=\epsfxsize \multiply\epsfysize\epsftmp
 \multiply\epsftmp\epsftsize \advance\epsfrsize-\epsftmp
 \epsftmp=\epsfxsize
 \loop \advance\epsfrsize\epsfrsize \divide\epsftmp 2
 \ifnum\epsftmp>0
 \ifnum\epsfrsize<\epsftsize\else
 \advance\epsfrsize-\epsftsize \advance\epsfysize\epsftmp \fi
 \repeat
 \epsfrsize=0pt
 \else
 \epsfrsize=\epsfysize
 \fi
 \fi
%
%
 \ifepsfverbose\message{#1: width=\the\epsfxsize, height=\the\epsfysize}\fi
 \epsftmp=10\epsfxsize \divide\epsftmp\pspoints
 \vbox to\epsfysize{\vfil\hbox to\epsfxsize{%
 \ifnum\epsfrsize=0\relax
 \includegraphics{#1}%
 \else
 \epsfrsize=10\epsfysize \divide\epsfrsize\pspoints
 \includegraphics{#1}%
 \fi
 \hfil}}%
\global\epsfxsize=0pt\global\epsfysize=0pt}%
%
%
 {\catcode`\%=12 \global\let\epsfpercent=
%
%
\long\def\epsfaux#1#2:#3\\{\ifx#1\epsfpercent
 \def\testit{#2}\ifx\testit\epsfbblit
 \epsfgrab #3 . . . \\%
 \epsffileokfalse
 \global\epsfbbfoundtrue
 \fi\else\ifx#1\par\else\epsffileokfalse\fi\fi}%
%
%
\def\epsfempty{}%
\def\epsfgrab #1 #2 #3 #4 #5\\{%
\global\def\epsfllx{#1}\ifx\epsfllx\epsfempty
 \epsfgrab #2 #3 #4 #5 .\\\else
 \global\def\epsflly{#2}%
 \global\def\epsfurx{#3}\global\def\epsfury{#4}\fi}%
%
%
\def\epsfsize#1#2{\epsfxsize}
%
%
\let\epsffile=\epsfbox
\def\att{t \bar{t}}
\def\app{p \bar{p}}
\def\rts{\sqrt{s}}
\def\mt{m_t}
\def\mb{m_b}
\def\mc{m_c}
\newcommand{\bksgam}{\ $B \to K^*+ \gamma$}
\newcommand{\brogam}{\ $B \to \rho+ \gamma$}
\def\BDSl{B \to D^* \ell \nu_\ell}
\def\vdvp{v \cdot v^\prime}
\def\xiaoo{\xi_{A_1}(\vdvp =1 )}
\def\Vbc{V_{cb}}
\newcommand{\Tosc}{T_{osc}}
\newcommand{\sqrts}{\sqrt{s}}
\newcommand{\bg}{\beta \gamma}
\newcommand{\xds}{x_i}
\newcommand{\Ds}{D_s^\pm}
\newcommand{\bb}{B^0 B^0}
\newcommand{\barbar}{{\overline{B^0}}\thinspace{\overline{B^0}}}
\newcommand{\barb}{B^0 {\overline{B^0}}}
\newcommand{\bbar}{$B^0$-${\overline{B^0}}$}
\newcommand{\Deltat}{\Delta t}
\newcommand{\delt}{\delta t}
\newcommand{\delmd}{\Delta M_d}
\newcommand{\delms}{\Delta M_s}
\newcommand{\ps}{10^{-12} s}
\newcommand{\zbbar}{Z^0 \to b {\overline{b}}}
\newcommand{\eebbx}{$e^+ e^- \to B {\overline{B}} X$}
\newcommand{\pbpbbx}{$p{\overline{p}} \to B {\overline{B}} X$}
\newcommand{\kkbar}{$K^0$-${\overline{K^0}}$}
\newcommand{\bdbdbar}{$B_d^0$-${\overline{B_d^0}}$}
\newcommand{\bsbsbar}{$B_s^0$-${\overline{B_s^0}}$}
\newcommand{\as}{\mbox{$\alpha_{\displaystyle s}$}}
\newcommand{\aso}{\mbox{$O(\alpha_{\displaystyle s})$}}
\newcommand{\ass}{\mbox{$O(\alpha_{\displaystyle s}^2)$}}
\newcommand{\asq}{\mbox{$\alpha_{\displaystyle s}(Q^2)$}}
\newcommand{\cc}{\mbox{$c {\overline{c}}$}}
\newcommand{\qq}{\mbox{$q {\overline{q}}$}}
\newcommand{\jp}{\mbox{$J/\Psi$}}
\newcommand{\lqc}{\Lambda_{QCD}}
\newcommand{\pmi}{{\not{p}}_{\perp}}
\newcommand{\set}{\sum E_{\perp}}
\newcommand{\ptr}{p_{\perp}}
\newcommand{\sww}{\sin^2{\theta_W}}
\newcommand{\sw}{\sin{\theta_W}}
 
 
\begin{flushright}
DESY 99-049\\
April 1999\\
\end{flushright}
\vspace*{1.5cm}
\begin{center} 
{\Large \bf
\centerline{CP Violation and Prospects at B Factories and Hadron
Colliders}}
 \vspace*{1.5cm}
 {\large A.~Ali}
\vskip0.2cm
 Deutsches Elektronen-Synchrotron DESY, Hamburg \\
Notkestra\ss e 85, D-22603 Hamburg, FRG\\

\vspace*{8.0cm}
{\large
Invited Talk; To be published in the Proceedings of the
13th Topical\\ Conference on Hadron Collider Physics,
TIFR, Mumbai, India,\\ January 14 - 20, 1999}
 
\end{center}]
  
\newpage

\title{CP Violation and Prospects at B Factories and Hadron 
Colliders}

\author{Ahmed Ali}

\address{Deutsches Elektronen Synchrotron DESY, Notkestra\ss e 85, 
D-22603 Hamburg, FRG\\E-mail: ali@x4u2.desy.de}

\maketitle

\abstracts{
We review the information on the CKM matrix elements, unitarity triangle
and CP-violating phases $\alpha$, $\beta$ and $\gamma$ in the standard 
model which will be measured in the forthcoming experiments at B 
factories, HERA-B and hadron colliders. We also discuss  
two-body non-leptonic decays $B \to h_1 h_2$, with $h_i$ being 
light mesons, which are interesting
from the point of view of CP violation and measurements of these 
phases. Partial rate CP asymmetries are presented in a number of decay 
modes using factorization for the matrix elements of the operators
in the effective weak Hamiltonian. 
Estimates of the branching ratios in this 
framework are compared with existing data on $B \to K\pi,\eta^\prime K, 
K^{*}\pi,\rho\pi$ decays from the CLEO collaboration.}
\section{Introduction}

We shall review the following three topics in quark flavour physics:
\begin{itemize}
\item An update of the Cabibbo-Kobayashi-Maskawa CKM matrix 
\cite{CKM}.
\end{itemize}
Here, the results of a global fit of the CKM parameters  yielding  
present profiles of the
unitarity triangle and CP-violating phases $\alpha$, $\beta$ and $\gamma$
and their correlations in the standard model (SM) are summarized 
\cite{AL99-1}.
\begin{itemize}
\item Estimates of the CP-violating partial rate asymmetries for charmless 
non-leptonic decays
$B \to h_1 h_2$, where $h_1$ and $h_2$ are light mesons, based on 
next-to-leading-order perturbative QCD and the factorization approximation
in calculating the matrix elements of the operators in the effective 
Hamiltonian approach \cite{AKL98-2}.
 \end{itemize}
Here, we first discuss a general classification of the CP-violating 
asymmetries in non-leptonic $B$ decays and then give updated numerical 
estimates for a fairly large number of two-body decays involving 
penguin- and tree-transitions \cite{AKL98-2}. Most of the decays considered 
here have branching 
ratios which are estimated to be in excess of $10^{-6}$ (and some in 
excess of $10^{-5}$) and many have measurable CP asymmetries. Hence, 
they are of interest for experiments at $B$ factories and hadron machines.  
\begin{itemize} 
\item Comparison of the branching ratios for $B \to h_1 h_2$ decays 
measured by the CLEO collaboration \cite{Gao-Wuerthwein-99,lss98} with the 
factorization-based theoretical estimates of the same 
\cite{AG97,ACGK97,AKL98-1}.
 \end{itemize}
The interest in these decays lies in that they
provide first information on the QCD penguin-amplitudes and the
CKM-suppressed non-leptonic $b \to u$ transitions. Hence, they will 
provide complementary information on the CKM matrix elements. It is argued 
that present data supports the factorization approach though it is not 
conclusive.  
 
\section{SM Fits of the CKM Parameters and the CP-Violating Phases 
$\alpha$, $\beta$ and $\gamma$}
Within the standard model (SM), CP violation is due to the presence of
a nonzero complex phase in the Cabibbo-Kobayashi-Maskawa (CKM) quark
mixing matrix $V$. We shall use the parametrization of
the CKM matrix due to Wolfenstein \cite{Wolfenstein}:
\beq
V \simeq \left(\matrix{
 1-{1\over 2}\lambda^2 & \lambda
 & A\lambda^3 \left( \rho - i\eta \right) \cr
 -\lambda ( 1 + i A^2 \lambda^4 \eta )
& 1-{1\over 2}\lambda^2 & A\lambda^2 \cr
 A\lambda^3\left(1 - \rho - i \eta\right) & -A\lambda^2 & 1 \cr}\right)~,
\label{CKM}
\eeq
which  has four {\it a priori} unknown parameters $A$, $\lambda$, $\rho$ and
$\eta$, where $\lambda$ is the Cabibbo angle and $\eta$ represents the
Kobayashi-Maskawa phase.
The allowed region in $\rho$--$\eta$ space can be elegantly displayed
using the so-called unitarity triangle (UT). While one has six 
such relations, 
resulting from the unitarity of the CKM matrix, the one written below has 
received particular attention: %
\beq
V_{ud} V_{ub}^* + V_{cd} V_{cb}^* + V_{td} V_{tb}^* = 0~.
\eeq
Using the form of the CKM matrix in Eq.~(\ref{CKM}), this can be recast as
\beq
\label{trianglerel}
\frac{V_{ub}^*}{\lambda V_{cb}} + \frac{V_{td}}{\lambda V_{cb}} = 1~,
\eeq
which is a triangle relation in the complex plane (i.e.\ $\rho$--$\eta$
space). Thus, allowed values of $\rho$
and $\eta$ translate into allowed shapes of the unitarity triangle.

 The interior CP-violating angles $\alpha$,
$\beta$ and $\gamma$ can be measured through CP asymmetries in $B$
decays. Likewise, some of these these angles can also be measured through
the decay rates. Additional constraints come from CP violation
in the kaon system ($\abseps$), as well as \bsbsbar\ mixing. In future,
the decays $B \to (X_s,X_d) \gamma$, $B \to (X_s,X_d) \ell^+ \ell^-$
and $K \to \pi \nu \bar{\nu}$ will further constrain the CKM matrix.

\subsection{Input Data}
The experimental and theoretical data which presently constrain the CKM
parameters $\lambda$, $A$, $\rho$ and $\eta$ are summarized below.

\begin{itemize}
\item $\vert V_{us}\vert$, $\vert V_{cb}\vert$ and $\vert 
V_{ub}/V_{cb}\vert$:
\end{itemize}
 We recall that 
$\vert V_{us}\vert$ has been extracted with good accuracy from $K\to\pi
  e\nu$ and hyperon decays \cite{PDG98} to be
$\vert V_{us}\vert=\lambda=0.2196\pm 0.0023$.
The determination of $\absvcb$ is based on the combined analysis of the
  inclusive and exclusive $B$ decays \cite{PDG98}: 
$ \vert V_{cb} \vert = 0.0395 \pm 0.0017$, yielding
$A = 0.819 \pm 0.035$. The knowledge of the CKM matrix
  element ratio $|V_{ub}/V_{cb}|$ is based on the analysis of the
  end-point lepton energy spectrum in semileptonic decays $B \to X_{u}
  \ell \nu_\ell$ and the measurement of the exclusive semileptonic
  decays $B \to (\pi, \rho) \ell \nu_\ell$. Present measurements in
  both the inclusive and exclusive modes are compatible with
  \cite{Parodiconf98}:
$\left\vert \frac{V_{ub}}{V_{cb}} \right\vert = 0.093\pm 0.014$.
This gives $\sqrt{\rho^2 + \eta^2} = 0.423 \pm 0.064$.
\begin{itemize}
\item {$ \abseps, \hat{B}_K$}:
\end{itemize}
 The experimental value of
$\abseps$ is \cite{PDG98}:
\beq
\abseps = (2.280\pm 0.013)\times 10^{-3}~.
\eeq
In the standard model, $\abseps$ is essentially proportional to the
imaginary part of the box diagram for \kkbar\ mixing and is given by
\cite{Burasetal}
\begin{eqnarray}
\abseps &=& \frac{G_F^2f_K^2M_KM_W^2}{6\sqrt{2}\pi^2\Delta M_K}
\hat{B}_K\left(A^2\lambda^6\eta\right)
\bigl(y_c\left\{\hat{\eta}_{ct}f_3(y_c,y_t)-\hat{\eta}_{cc}\right\}
 \nonumber \\
&~& ~~~~~~~~~~~~~~+ 
~\hat{\eta}_{tt}y_tf_2(y_t)A^2\lambda^4(1-\rho)\bigr), 
\label{eps}
\end{eqnarray}
where $y_i\equiv m_i^2/M_W^2$, and the functions $f_2$ and $f_3$
are the Inami-Lim function \cite{InamiLim}.
Here, the $\hat{\eta}_i$ are QCD correction factors, calculated at
next-to-leading order: ($\hat{\eta}_{cc}$) \cite{HN94},
($\hat{\eta}_{tt}$) \cite{etaB} and ($\hat{\eta}_{ct}$) \cite{HN95}.
The theoretical uncertainty in the expression for $\abseps$ is in the
renormalization-scale independent parameter $\hat{B}_K$, which
represents our ignorance of the hadronic matrix element $\langle K^0
\vert {({\overline{d}}\gamma^\mu (1-\gamma_5)s)}^2 \vert
{\overline{K^0}}\rangle$. Recent calculations of $\hat{B}_K$ using
lattice QCD methods are summarized at the 1998 summer conferences by Draper 
\cite{Draper98} and Sharpe \cite{Sharpe98}, yielding
\begin{equation}
 \hat{B}_K=0.94 \pm 0.15 .
\label{BKrange}
\end{equation}
\begin{itemize}
\item {$ \Delta M_d, \fbb$}:
\end{itemize}
 The present world average for
  $\Delta M_d$ is \cite{Alexander98}
\beq
\Delta M_d = 0.471 \pm 0.016~(ps)^{-1} ~.
\eeq
The mass difference $\Delta M_d$ is calculated from the \bdbdbar\ box
diagram, dominated by $t$-quark exchange:
\beq
\label{bdmixing}
\Delta M_d = \frac{G_F^2}{6\pi^2}M_W^2M_B\left(\fbb\right)\hat{\eta}_B y_t
f_2(y_t) \vert V_{td}^*V_{tb}\vert^2~, \label{xd}
\eeq
where, using Eq.~(\ref{CKM}), $\vert V_{td}^*V_{tb}\vert^2=
A^2\lambda^{6} [\left(1-\rho\right)^2+\eta^2]$. Here, $\hat{\eta}_B$
is the QCD correction, which has 
the value $\hat{\eta}_B=0.55$, calculated in the $\overline{MS}$
scheme \cite{etaB}.

For the $B$ system, the hadronic uncertainty is given by $\fbb$.
 Present estimates of this quantity using lattice QCD yield $\fbd
\sqrt{\hat{B}_{B_d}} =(190 \pm 23)$ MeV in the quenched approximation
\cite{Draper98,Sharpe98}.
The effect of unquenching is not yet understood completely. Taking the
MILC collaboration estimates of unquenching would increase the central
value of $\fbd \sqrt{\hat{B}_{B_d}}$ by $21$ MeV \cite{MILC98}. In the
fits discussed here \cite{AL99-1}, the following range has been used
\begin{equation}
\fbd \sqrt{\hat{B}_{B_d}} = 215 \pm 40 ~\mbox{MeV}~.
\label{FBrange}
\end{equation}
\begin{itemize}
\item {$ \Delta M_s, \fbbs$}:
\end{itemize}
 The \bsbsbar\ box
  diagram is again dominated by $t$-quark exchange, and the mass
  difference between the mass eigenstates $\delms$ is given by a
  formula analogous to that of Eq.~(\ref{xd}):
\beq
\delms = \frac{G_F^2}{6\pi^2}M_W^2M_{B_s}\left(\fbbs\right)
\hat{\eta}_{B_s} y_t f_2(y_t) \vert V_{ts}^*V_{tb}\vert^2~.
\label{xs}
\eeq
Using the fact that $\vert V_{cb}\vert=\vert V_{ts}\vert$ (Eq.~(\ref{CKM})),
it is clear that one of the sides of the unitarity triangle, $\vert
V_{td}/\lambda V_{cb}\vert$, can be obtained from the ratio of $\delmd$ and
$\delms$,
\beq
\frac{\delms}{\delmd} =
 \frac{\hat{\eta}_{B_s}M_{B_s}\left(\fbbs\right)}
{\hat{\eta}_{B_d}M_{B_d}\left(\fbb\right)}
\left\vert \frac{V_{ts}}{V_{td}} \right\vert^2.
\label{xratio}
\eeq
The only real uncertainty in this quantity is the ratio of hadronic
matrix elements $\fbbs/\fbb$. Present estimate of this quantity is
\cite{Draper98,Sharpe98}:
\beq
\label{xirange}
\xi_s=1.14\pm 0.06 ~.
\eeq
The present lower bound on $\Delta M_s$ is: $\Delta M_s > 12.4
~\mbox{(ps)}^{-1}$ (at $95\%$ C.L.) \cite{Parodiconf98}.

\begin{table}
\caption{Data used in the CKM fits.}
\label{datatable}
\hfil
\vbox{\offinterlineskip
\halign{&\vrule#&
 \strut\quad#\hfil\quad\cr
\noalign{\hrule}
height2pt&\omit&&\omit&\cr
& Parameter && Value & \cr
height2pt&\omit&&\omit&\cr
\noalign{\hrule}
height2pt&\omit&&\omit&\cr
\footnotesize
& $\lambda$ && $0.2196$ & \cr
& $\vert V_{cb} \vert $ && $0.0395 \pm 0.0017$ & \cr
& $\vert V_{ub} / V_{cb} \vert$ && $0.093 \pm 0.014$ & \cr
& $\abseps$ && $(2.280 \pm 0.013) \times 10^{-3}$ & \cr
& $\Delta M_d$ && $(0.471 \pm 0.016)~(ps)^{-1}$ & \cr
& $\Delta M_s$ && $ > 12.4 ~(ps)^{-1}$ & \cr 
& $\overline{\mt}(\mt(pole))$ && $(165 \pm 5)$ GeV & \cr
& $\overline{\mc}(\mc(pole))$ && $1.25 \pm 0.05$ GeV & \cr
& $\hat{\eta}_B$ && $0.55$ & \cr
& $\hat{\eta}_{cc} $ && $1.38 \pm 0.53$ & \cr
& $\hat{\eta}_{ct} $ && $0.47 \pm 0.04$ & \cr
& $\hat{\eta}_{tt} $ && $0.57$ & \cr
& $\hat{B}_K$ && $0.94 \pm 0.15$ & \cr
& $\fbd\sqrt{\hat{B}_{B_d}} $ && $215 \pm 40$ MeV & \cr
& $\xi_s $ && $1.14 \pm 0.06$  & \cr
height2pt&\omit&&\omit&\cr
\noalign{\hrule}}}
\end{table}

There are two other measurements which should be mentioned here.
First, the KTEV collaboration \cite{KTEV99} has recently reported a
measurement of direct CP violation in the $K$ sector through the ratio
$\epsilon^\prime/\epsilon$, with
\beq
{\rm Re} (\epsilon^\prime/\epsilon) = \left( 28.0 \pm 3.0 (\mbox{stat}) 
\pm 2.6 (\mbox{syst}) \pm 1.0(\mbox{MC stat})\right) \times 10^{-4} ~,
\eeq
in agreement with the earlier measurement by the CERN experiment NA31
\cite{NA31}, which reported a value of $(23 \pm 6.5) \times 10^{-4}$
for the same quantity. The present world average is ${\rm Re}
(\epsilon^\prime/\epsilon) =(21.8 \pm 3.0) \times 10^{-4}$. This
combined result excludes the superweak model \cite{superweak} by more
than $7\sigma$. 

A great deal of theoretical effort has gone into calculating this
quantity at next-to-leading order accuracy in the SM
\cite{Buraseps,Martinellieps,Bertolinieps}. The result of this
calculation can be summarized in the following form due to Buras and
Silvestrini \cite{BS98}:
\beq
{\rm Re} (\epsilon^\prime/\epsilon) = {\rm Im} \lambda_t 
\left[ -1.35 + R_s \left( 1.1 \vert r_Z^{(8)} \vert B_6^{(1/2)} 
+ (1.0 -0.67 \vert r_Z^{(8)} \vert) B_8^{(3/2)} \right) \right]~.
\label{epsilonp}
\eeq
Here $\lambda_t= V_{td}V_{ts}^* = A^2 \lambda^4 \eta$ and $r_Z^{(8)}$
represents the short-distance contribution, which at the NLO precision
is estimated to lie in the range $ 6.5 \leq \vert r_Z^{(8)} \vert \leq
8.5$ \cite{Buraseps,Martinellieps}. The quantities
$B_6^{(1/2)}=B_6^{(1/2)}(m_c)$ and $B_8^{(3/2)}=B_8^{(3/2)}(m_c)$ are
the matrix elements of the $\Delta I=1/2$ and $\Delta I =3/2$
operators $O_6$ and $O_8$, respectively, calculated at the scale
$\mu=m_c$. Lattice-QCD \cite{B68-Latt} and the $1/N_c$ expansion
\cite{BurasNc} yield:
\beq
0.8 \leq B_6^{(1/2)} \leq 1.3,~~~~~~~~~~~0.6 \leq B_8^{(3/2)} \leq 1.0~.
\eeq
Finally, the quantity $R_s$ in Eq.~(\ref{epsilonp}) is defined as:
\beq
R_s \equiv \left( \frac{150~\mbox{MeV}}{m_s(m_c) + m_d(m_c)} \right)^2 ~,
\label{Rsdef}
\eeq
essentially reflecting the $s$-quark mass dependence. The present
uncertainty on the CKM matrix element is $\pm 23\%$, which is already
substantial. However, the theoretical uncertainties related to the
other quantities discussed above are considerably larger. For
example, the ranges $\epsilon^\prime/\epsilon=(5.3 \pm 3.8) \times
10^{-4}$ and $\epsilon^\prime/\epsilon=(8.5 \pm 5.9) \times 10^{-4}$,
assuming $m_s(m_c)=150\pm 20$ MeV and $m_s(m_c)=125\pm 20$ MeV,
respectively, have been quoted as the best representation of the
status of $\epsilon^\prime/\epsilon$ in the SM \cite{Burasreview98}.
These estimates are somewhat on the lower side compared to the data
but not inconsistent.

Thus, whereas $\epsilon^\prime/\epsilon$ represents a landmark
measurement, establishing for the first time direct CP-violation 
in decay amplitudes, and  hence removing the superweak model of 
Wolfenstein and its  various incarnations from further consideration, its 
impact on CKM phenomenology,
particularly in constraining the CKM parameters, is marginal. For
this reason, the measurement of
$\epsilon^\prime/\epsilon$ is not included in the CKM fits 
summarized here.

Second, the CDF collaboration has recently made a measurement of $\sin
2\beta$ \cite{CDF99,Bauer-99}. In the Wolfenstein parametrization, 
$-\beta$ is
the phase of the CKM matrix element $V_{td}$. From Eq.~(\ref{CKM}) one
can readily find that
\beq
\sin (2 \beta) = \frac{2\eta(1-\rho)}{(1-\rho)^2 + \eta^2} ~.
\eeq
Thus, a measurement of $\sin 2\beta$ would put a strong contraint on
the parameters $\rho$ and $\eta$. However, the CDF measurement gives
\cite{CDF99}
\beq
\sin 2\beta = 0.79^{+0.41}_{-0.44} ~,
\eeq
or $\sin 2\beta > 0$ at 93\% C.L. 
This constraint is quite weak -- the indirect measurements already
constrain $0.52 \le \sin 2\beta \le 0.94$ at the 95\% C.L.\ in the SM
\cite{AL99-1}.
(The CKM fits reported recently in the 
literature \cite{Parodiconf98,Mele98,PRS98} yield
similar ranges.)  In light of this, this measurement is not included in 
the fits. The data used in the CKM fits are summarized in Table 
\ref{datatable}.
\subsection{SM Fits}

 In
the fit presented here \cite{AL99-1}, ten parameters are allowed to vary: 
$\rho$, $\eta$, $A$, 
$m_t$, $m_c$, $\eta_{cc}$, $\eta_{ct}$, $f_{B_d} \sqrt{\hat{B}_{B_d}}$,
$\hat{B}_K$, and $\xi_s$. The $\Delta M_s$ constraint is included using
the amplitude method \cite{Moser97}. The rest of the parameters are fixed to 
their central values.  
The allowed (95\% C.L.) $\rho$--$\eta$ region is shown in
Fig.~\ref{rhoeta1}. The best fit has $(\rho,\eta) = (0.20,0.37)$.

\begin{figure}
\vskip -1.0truein
\centerline{\epsfxsize 3.5 truein \epsfbox {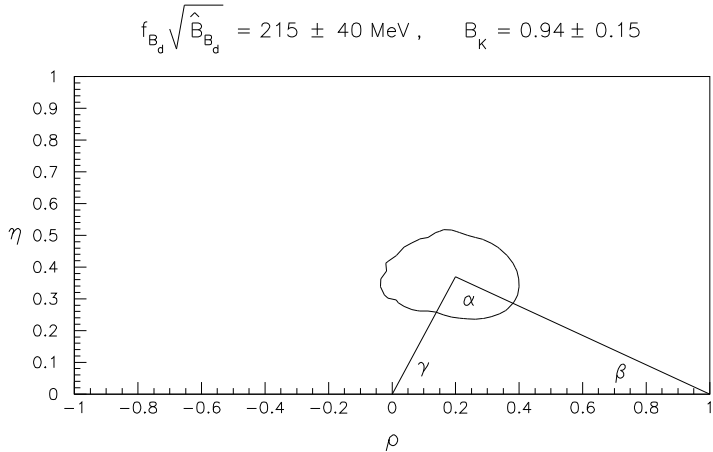}}
\vskip -1.4truein
\caption{Allowed region in $\rho$--$\eta$ space in the SM, from a
  fit to the ten parameters discussed in the text and given in Table
  \protect{\ref{datatable}}. The limit on $\Delta M_s$ is included
  using the amplitude method \protect\cite{Moser97}. The
  theoretical errors on $\fbd\protect\sqrt{\hat{B}_{B_d}}$,
  $\hat{B}_K$ and $\xi_s$ are treated as Gaussian. The solid line
  represents the region with $\chi^2=\chi_{min}^2+6$ corresponding to
  the 95\% C.L.\ region. The triangle shows the best fit.
(From Ref.~2.)}
\label{rhoeta1}
\end{figure}

The CP angles $\alpha$, $\beta$ and $\gamma$ can be measured in
CP-violating rate asymmetries in $B$ decays. These angles can be
expressed in terms of $\rho$ and $\eta$. Thus, different shapes of the
unitarity triangle are equivalent to different values of the CP
angles. Referring to Fig.~\ref{rhoeta1}, we note that the preferred
(central) values of these angles are $(\alpha,\beta,\gamma) =
(93^\circ,25^\circ,62^\circ)$. The allowed ranges at 95\% C.L. are
\begin{eqnarray}
\label{CPangleregion}
65^\circ \le & \alpha & \le 123^\circ \nn\cr
16^\circ \le & \beta & \le 35^\circ \nn\cr
36^\circ \le & \gamma & \le 97^\circ
\end{eqnarray}

Of course, the values of $\alpha$, $\beta$ and $\gamma$ are
correlated, i.e.\ they are not all allowed simultaneously. 
 We illustrate these
correlations in Figs.~\ref{alphabetacorr} and \ref{alphagammacorr}.
Fig.~\ref{alphabetacorr} shows the allowed
region in $\sin 2\alpha$--$\sin 2\beta$ space allowed by the data. And
Fig.~\ref{alphagammacorr} shows the allowed (correlated) values of the
CP angles $\alpha$ and $\gamma$. This correlation is roughly linear,
due to the relatively small allowed range of $\beta$
(Eq.~(\ref{CPangleregion})).

 We remark that the correlations shown are 
specific to the SM and are expected to be different, in general, in non-SM
scenarios. A comparative study for some variants of the minimal 
supersymmetric model (MSSM) has been presented recently \cite{AL99-1}, 
underlying the importance of measuring the angles $\alpha$, $\beta$ and 
$\gamma$ precisely. One expects almost similar constraints on 
$\beta$ from the CKM fits in  the SM and MSSM, but $\alpha$ and $\gamma$ 
may provide a discrimination.

\begin{figure}
\vskip -1.0truein
\centerline{\epsfxsize 3.5 truein \epsfbox {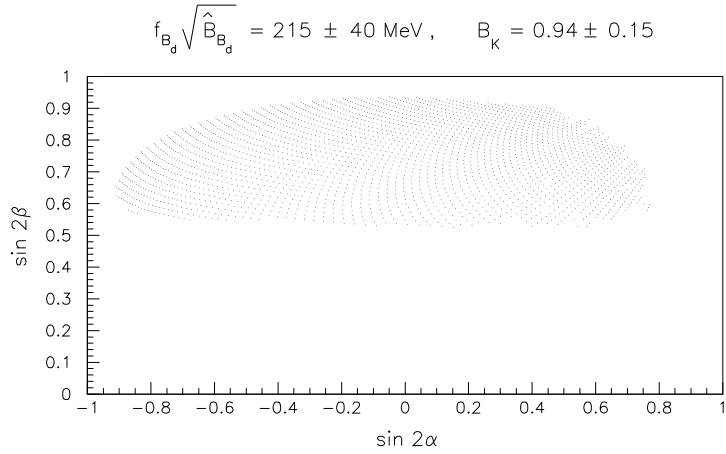}}
\vskip -1.4truein
\caption{Allowed 95\% C.L. region of the CP-violating quantities 
  $\sin 2\alpha$ and $\sin 2\beta$ in the SM, from a fit to the data given in
  Table \protect{\ref{datatable}}.}
\label{alphabetacorr}
\end{figure}

\begin{figure}
\vskip -1.0truein
\centerline{\epsfxsize 3.5 truein \epsfbox {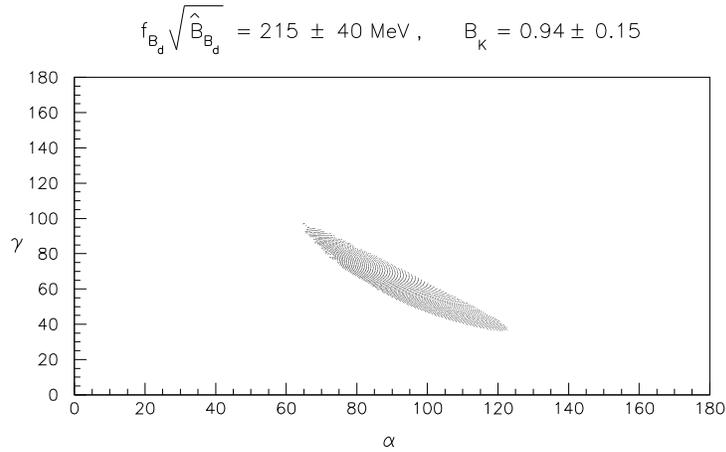}}
\vskip -1.4truein
\caption{Allowed 95\% C.L. region of the CP-violating quantities 
  $\alpha$ and $\gamma$ in the SM, from a fit to the data given in Table
  \protect{\ref{datatable}}.}
\label{alphagammacorr}
\end{figure}
\section{CP-Violating Asymmetries in $B^\pm \to (h_1 h_2)^\pm$ Decays}
Apart from the decay modes $B \to J/\psi K_s^0$ and $B \to \pi 
\pi$, discussed at great length in the literature, there are many 
interesting two-body 
decays $B \to h_1 h_2$  which are expected to
have large CP asymmetries in their partial decay rates.
Recent measurements by the CLEO collaboration 
of $B$-decays into final state such as $h_1h_2=
\pi K, \eta^\prime K, \pi \rho, \pi K^*$ have rekindled theoretical
interest in these decays \cite{Gao-Wuerthwein-99,lss98}. 
A completely quantitative description of these and related 
two-body decays is a challenging enterprise, as this requires
knowledge of the four-quark-operator matrix elements in the decays $B \to 
h_1 h_2$, for which the QCD technology is not yet ripe.
Hence, calculations of the decay amplitudes from first principles
in QCD are difficult and a certain amount of model-building is 
unavoidable. 
Here, we shall summarize the work done in estimating the rates
\cite{AG97,ACGK97,AKL98-1} and CP asymmetries \cite{AKL98-2} in some
selective decay modes, based on perturbative QCD and factorization.

For charged $B^\pm$ decays the CP-violating rate-asymmetries 
in partial decay rates are defined as follows:
\begin{equation}
  \label{acp}
  A_{CP} \equiv \frac{  \Gamma ( B^+ \to  f^+)-\Gamma ( B^- \to  f^-) }{
\Gamma ( B^+ \to  f^+)+  \Gamma ( B^- \to  f^-) }~,
\end{equation}
where $f^\pm =(h_1h_2)^\pm$. 
To be non-zero, these asymmetries require both weak and 
strong phase differences in interfering amplitudes. 
The weak phase difference arises from the superposition of amplitudes from 
the various tree- and penguin-diagrams, with the former involving $b \to u$ 
and the latter $b\to s$ or $b \to d$ transitions. The strong phases,
which are needed to obtain non-zero values for $A_{CP}$ in (\ref{acp}),
are generated by final state interactions. This is modeled using 
perturbative QCD by taking into account the NLO corrections,
following earlier suggestions along these lines \cite{SEW}. It should be
stressed that this formalism includes not only the so-called {\it charm 
penguins}
\cite{marti-penguin-97} but {\it all} penguins (as well as the 
tree-contribution) in the framework of an effective Hamiltonian.

\subsection{CP-violating Asymmetries Involving $b \to s$ Transitions}
For the $b \to s$, and the charge conjugated $\bar{b} \to 
\bar{s}$, transitions, the respective  decay amplitudes ${\cal M}$ and 
$\overline{\cal M}$,  
including the weak and strong phases, can be generically written as:
\begin{eqnarray}
{\cal M}&=&  T \xi _u -P_{tc} \xi_t  e^{i\delta_{tc}}
-P_{uc} \xi_u  e^{i\delta_{uc}},\nonumber\\
\overline{\cal M} &=&  T \xi^* _u -P_{tc} \xi_t^*  e^{i\delta_{tc}}
 -P_{uc} \xi_u^*  e^{i\delta_{uc}},\label{msim}
\end{eqnarray}
where we define
\begin{eqnarray}
P_{tc}  e^{i\delta_{tc}}&\equiv&  P_t e^{i\delta_t} -P_c e^{i\delta_c} 
,\nonumber\\
P_{uc}  e^{i\delta_{uc}}&\equiv&  P_u   e^{i\delta_u}-P_c e^{i\delta_c}.
\end{eqnarray}
Here $\xi_i=V_{ib}V^*_{is}$ and use has been made of the unitarity 
relation $\xi_c=-\xi_u-\xi_t $. In the above expressions $T$ denotes  
the contributions from the current-current operators;  $P_t$, $P_c$ 
and $P_u$ denote the contributions from penguin operators proportional to
the product of the CKM matrix elements $\xi_t$, $\xi_c$ and $\xi_u$, 
and the corresponding 
strong phases are denoted by $\delta_t$, $\delta_c$ and $\delta_u$, 
respectively.
The explicit expressions for the CP-violating asymmetry $A_{CP}$
are, in general, not very illuminating \cite{AKL98-2}.
However, as the amplitudes involve several small parameters\footnote{The
smallness of these quantities reflects the CKM-suppression and/or
QCD dynamics calculated in perturbation theory.}, 
 much simplified forms for $A^-$ and $A^+$, and  hence $A_{CP}$, can be 
obtained in specific decays by keeping only the leading terms 
\cite{AKL98-1,AKL98-2}. 

To exemplify this, we note that  $|\xi_u| \ll | \xi_t|\simeq
|\xi_c|$, with an upper bound  $|\xi_u|/| \xi_t| \leq 0.025$.
In some channels, such as $B^\pm \to K^\pm \pi^0$, $K^{*\pm} \pi^0$,
$K^{*\pm} \rho^0$, typical
value of the ratio $|P_{tc}/T|$ is of $O(0.1)$, with both $P_{uc}$ and $P_{tc}$
comparable with typically $|P_{uc}/P_{tc}| =O(0.3)$.
 Using these approximations, 
the CP-violating asymmetry in $b \to s$ transitions can be expressed as  
\begin{equation}
  \label{cps1}
  A_{CP} \simeq \frac{2z_{12} \sin \delta_{tc} \sin \gamma}
{1+ 2 z_{2} \cos \delta_{tc} \cos \gamma +z_{2}^2},\label{app1}
\end{equation}
where $z_{2}=|\xi_u/\xi_t|\times T/P_{tc}$. Note that $A_{CP}$ is 
approximately proportional to $\sin \gamma$, as pointed out by
Fleischer and Mannel \cite{FM97} in the context of the decay $B \to K\pi$.
Due to the circumstance that the suppression due to $|\xi_u/\xi_t|$ is
 stronger than the enhancement
due to $T/P_{tc}$, restricting the value of $z_{2}$,  the CP-violating 
asymmetry for these kinds of decays are expected to be O(10\%). Explicit
calculations in model estimates confirm this pattern \cite{AKL98-2}.

There are also decay modes with vanishing tree contributions, such as 
$B^\pm \to \pi^\pm K_S^0$, $\pi^\pm K^{*0}$, $\rho^\pm K^{*0}$. With 
$T=0$ and $|\xi_{u}|\ll  |\xi_{t}|$, 
the CP-violating asymmetry can now be expressed as 
\begin{equation}
  \label{cps2}
  A_{CP} \simeq 2\frac{P_{uc}}{P_{tc}}\left |\frac{\xi_u}{\xi_t}\right |
 \sin (\delta_{uc}-\delta_{tc}) \sin \gamma.
\end{equation}
As $P_{uc}/P_{tc}\ll 1$, and also
$|\xi_u/\xi_t| \ll 1$, the   
CP-violating asymmetries are expected to be small. 
Some representative estimates are \cite{AKL98-2}: 
$A_{CP}(\pi^\pm K_s^0)=-1.5\%$, $A_{CP}(\pi^\pm \optbar{K^{*0}})=-1.7\%$,
$A_{CP}(\rho^\pm \optbar{K^{*0}})=-1.7\%$. In scenarios with additional
CP-violating phases, these CP asymmetries can be greatly enhanced and 
hence they are of interest in searching for non-SM CP-violation effects 
in $B$ decays.

\subsection{CP-violating Asymmetries Involving $b \to d$ Transitions}

For  $b\to d$ transitions, the decay amplitudes can be expressed as
\begin{eqnarray}
{\cal M} &=&  T \zeta _u -P_{tc} \zeta_t  e^{i\delta_{tc}}
 -P_{uc} \zeta_u  e^{i\delta_{uc}},\nonumber\\
\overline{\cal M} &=&  T \zeta^* _u -P_{tc} \zeta_t^*  e^{i\delta_{tc}}
 -P_{uc} \zeta_u^*  e^{i\delta_{uc}},
\end{eqnarray}
where $\zeta_i=V_{ib}V^*_{id}$, and again the CKM unitarity has been
used in the form 
$\zeta_c = -\zeta_t-\zeta_u $.
For the tree-dominated decays involving $b\to d $ transitions, such as 
$B^\pm \to \pi^\pm \eta ^{(\prime)}$, $\rho^\pm \eta ^{(\prime)}$, $\rho^\pm 
\omega$, the relation $P_{uc} < P_{tc} \ll T$ holds, 
and the CP-violating asymmetry is approximately given by
\begin{equation}
 A_{CP} \simeq \frac{-2z_{1} \sin \delta_{tc} \sin \alpha}
{1+ 2 z_{1} \cos \delta_{tc} \cos \alpha},\label{app3}
\end{equation}
with $z_{1}=|\zeta_t/\zeta_u|\times TP_{tc}/T'^2$ and $T'^2\equiv T^2  - 
2 TP_{uc}\cos \delta_{uc}$. Note, the CP-violating asymmetry is 
approximately proportional to $\sin\alpha$ in this 
case. Concerning $z_1$, we note that
the suppression due to $P_{tc}T/T^{\prime 2} \ll 1$ is accompanied with some 
enhancement from
$|\zeta_t/\zeta_u|$ (the central value of this quantity is about 2.1 
\cite{AL99-1}),
 making the CP-violating asymmetry in this  kind of decays to have a
value  $A_{CP}=(5$-$10)\%$.

For the decays with vanishing tree contribution, such as 
$B^\pm \to K^\pm  K_S^0$, $K^\pm \optbar{K^{*0}}$,
$K^{*\pm} \optbar{K^{*0}}$, 
the CP-violating asymmetry is approximately  proportional to  $\sin\alpha$
again,
\begin{equation}
 A_{CP} = \frac{-2z_{3} \sin (\delta_{uc}-\delta_{tc}) \sin \alpha}
{1- 2 z_{3} \cos (\delta_{uc}-\delta_{tc}) \cos \alpha+z_3^2},\label{app4}
\end{equation}
with $z_3=|\zeta_u/\zeta_t|\times P_{uc}/P_{tc}$.
As the  suppression from  
$|\zeta_u/\zeta_t|$ and $|P_{uc}/P_{tc}|$  
is not very strong, the CP-violating asymmetry are typically 
of order $(10$-$20)\%$.

 We list the estimated CP asymmetries and branching ratios (charge-conjugate 
averaged) in 
$B^\pm \to (h_1 h_2)^\pm$ decays in the upper half of Table 
2, keeping only those decays which are expected to have branching 
ratios in excess of $10^{-6}$. While the listed $A_{CP}$ are not sensitive 
to the
precise values of the form factors, the branching ratios are; the numbers
given correspond to the BSW model \cite{BSW85}. We have indicated the
uncertainty on $A_{CP}$ resulting from the virtuality of the gluon $g(k^2) 
\to q_i \bar{q}_i$, influencing the absorptive parts of the amplitudes, for 
$k^2=m_b^2/2 \pm 2$ GeV$^2$.
\begin{table}[t]
\caption{CP-rate asymmetries $A_{CP}$ and charge-conjugate-averaged
branching ratios for some
selected $B  \to h_1 h_2$ decays, estimated in the factorization approach
\protect\cite{AKL98-2}, updated for the central values
of the CKM fits \protect\cite{AL99-1} $\rho=0.20$, $\eta=0.37$   
and the factorization model parameters $\xi=0.5$ and $k^2=m_b^2/2\pm 2$
GeV$^2$. \label{tab:cp}}
\begin{center}
\footnotesize
\begin{tabular}{|c|c|c|l|}
\hline
Decay Modes & \raisebox{0pt}[13pt][7pt]{CP-class}&
\raisebox{0pt}[13pt][7pt]{$A_{CP}
(\%)$}&\raisebox{0pt}[13pt][7pt]{$BR(\times 10^{-6})$}\\
\hline
$B^\pm \to K^\pm \pi^0$ &$ (i)$    &$ -7.7^{-2.2}_{+4.0}$ & $10.0$\\
\hline
$B^\pm \to K^{*\pm} \pi^0$ &$ (i)$ &$-14.4^{-4.4}_{+8.2}$ &$4.3$\\
\hline
$B^\pm \to K^{*\pm} \rho^0$ &$ (i)$ &$ -13.5^{-4.0}_{+7.5}$ &$4.8$\\
\hline
$B^\pm \to \eta \pi^\pm $ &$ (i)$   &$ 9.3^{+1.9}_{-4.1}$ & $5.5$ \\  
\hline
$B^\pm \to \eta^\prime \pi^\pm $&$ (i)$   &$ 9.4^{+2.1}_{-4.5}$ & $3.7$\\
\hline
$B^\pm \to \eta \rho^\pm $ &$ (i)$ &$3.1^{+0.7}_{-1.7}$ &$8.6$\\
\hline
$B^\pm \to \eta^\prime \rho^\pm $ &$ (i)$  &$3.1^{+0.7}_{-1.8}$ &$6.2$\\
\hline
$B^\pm \to \rho^\pm \omega$ & $ (i)$ & $7.0^{+1.5}_{-3.4}$ & $ 21.0$\\
\hline
$B^\pm \to \eta^\prime K^\pm$& $ (i)$ &$-4.9^{-1.2}_{+2.1} $ & $23.0$\\
\hline
$B^\pm \to \pi^\pm \rho^0 $& $ (i)$ &$ -3.1^{-0.9}_{+2.0}$ & $9.0$\\
\hline
$B^\pm \to \eta K^\pm  $&$ (i)$   &$ 7.0^{+2.9}_{-5.2}$ &$ 2.6$\\
\hline
$B^\pm \to \eta K^{*\pm} $ &$ (i)$ &$-10.5^{-3.1}_{+5.9}$ &$ 2.1$\\
\hline
$B^\pm \to K^\pm \omega$ &$ (i)$ &$-14.4^{-4.5}_{+8.1}$ &$3.2$\\
\hline
$B^\pm \to \pi^\pm \omega$ &$ (i)$ &$7.7^{+1.7}_{-3.7}$ &$9.5$\\
\hline
$B^\pm \to K^{*\pm} \omega$ &$ (i)$ &$ -10.3^{-3.1}_{+5.6}$ &$11.0$\\
\hline
$\optbar{B^0} \to K^{*\pm} \rho^\mp$ &$ (i)$ & $ -17.2^{-5.5}_{+9.8}$
&$5.4$\\
\hline
$\optbar{B^0} \to K^\pm \pi^\mp$ & $(i)$   &$ -8.2^{-2.3}_{+4.3}$ &$14.0$\\
\hline
$\optbar{B^0} \to \pi^\pm K^{*\mp}$ & $ (i)$ & $ -17.2^{-5.5}_{+9.8}$
& $ 6.0$\\
\hline
$\optbar{B^0} \to \eta^\prime K_S^0 $ & $(ii)$ &$ 33.6^{-0.2}_{+0.3} $
& $ 23.0$\\
 \hline
$\optbar{B^0} \to \pi^+\pi^-$ & $ (ii)$   &$ 25.4^{+0.2}_{-1.0}$ &$13.0$ \\
\hline
$\optbar{B^0} \to \pi^0\pi^0$ & $ (ii)$   &$ -45.1^{-1.8}_{+6.4}$ & $0.4$\\
\hline
$\optbar{B^0} \to K_S^0 \pi^0$ &$ (ii)$ & $ 39.6^{+0.5}_{-0.9}$ 
&$3.0$\\   
\hline
$\optbar{B^0} \to K_S^0 \eta$ &$ (ii)$ & $ 41.2^{+0.7}_{-1.1}$ &$1.0$\\
\hline
$\optbar{B^0} \to K_S^0 \phi$ &$ (ii)$ & $ 35.2$ &$9.0$\\
\hline
$\optbar{B^0} \to \rho^+ \rho^-$ &$ (iii)$ & $ 17.1^{+0.1}_{-0.6}$
&$24.0$\\
\hline
$\optbar{B^0} \to \rho^0 \rho^0$ &$ (iii)$ & $ -46.0^{-1.4}_{+4.3}$
&$1.0$\\
\hline
$\optbar{B^0} \to \omega \omega$ &$ (iii)$ & $ 56.5^{+1.3}_{-2.8}$
&$1.1$\\
\hline  
$B^0/\bar{B}^0 \to \rho^-\pi^+/\rho^+\pi^-$ & $(iv)$ &$ 
13.9^{-0.5}_{+1.0}$ & $7.8$\\
 \hline
$B^0/\bar{B}^0 \to \rho^+\pi^-/\rho^-\pi^+$ & $(iv)$ & $ 
9.7^{-0.6}_{+0.9}$ & $ 29.0$\\
 \hline
   \end{tabular} \end{center} 
\end{table}  
\section{CP-violating Asymmetries in $B^0 \to (h_1h_2)^0$ Decays}

The CP asymmetries involving the neutral $B^0(\bar{B}^0)$ decays may require 
time-dependent measurements.
Defining the time-dependent asymmetries as
\begin{equation}
  \label{ae}
  A_{CP}(t) \equiv \frac{ \Gamma (B^0(t) \to f) -\Gamma (\overline  B^0(t) 
\to  \bar f)}{
 \Gamma (B^0(t) \to f) +\Gamma ( \overline  B^0(t) \to  \bar  f)},
\end{equation}
there are four cases that one encounters for neutral $B^0(\bar{B}^0)$ 
decays:
 \begin{itemize}
\item case (i): $B^0 \to f $, $\bar{B}^0 \to \bar{f}$, where $f$ or $\bar{f}$
is not a common final state
of $B^0$ and $\bar{B}^0$, for example ${B}^0 \to K^+\pi^-$ and 
$\bar{B}^0 \to K^-\pi^+$ .
\item case (ii): $B^0 \to (f=\bar{f}) \leftarrow \bar{B}^0$ with 
$f^{CP}=\pm f$, involving final
states which are CP eigenstates, i.e., decays such as $\bar{B}^0 (B^0) \to 
\pi^+ \pi^-, \pi^0 \pi^0, K_S^0 \pi^0$ etc.
\item case (iii): $B^0 \to (f=\bar{f}) \leftarrow \bar{B}^0$, with
$f$ involving final states which are not CP eigenstates. They include
decays into two vector mesons $B^0 \to (VV)^0$, as the $VV$ states are not 
CP-eigenstates.  \item case (iv): $B^0 \to (f \& \bar{f}) \leftarrow 
\bar{B}^0$ with $f^{CP} \neq f$, i.e.,  both $f$ and  $\bar{f}$ are common 
final states of $B^0$ and $\bar{B}^0$, but they are not  CP eigenstates.
 Decays ${B}^0/\bar{B}^0  \to \rho^+ \pi^-$, $\rho^- \pi^+$ and 
$B^0/\bar{B}^0 \to K^{*0} 
 K_S^0$, $\bar K^{*0} K_S^0$ are two interesting examples in this class.
\end{itemize}
Here case (i) is very similar to the charged $B^\pm$ decays, discussed 
above. For case (ii),
and (iii), $A_{CP}(t)$ would involve $B^0$ - $\overline{B^0}$ mixing.
Assuming $|\Delta \Gamma | \ll |\Delta m |$
and $|\Delta \Gamma/\Gamma | \ll 1$, which hold in the standard model for
the mass and width differences $\Delta m $ and
$\Delta \Gamma$ in the neutral $B$-sector, one can express
$A_{CP}(t)$ in a simplified form:
\begin{equation}
\label{acpeps}
A_{CP}(t) \simeq a_{\epsilon^\prime} \cos (\Delta m t) + a_{\epsilon + 
\epsilon^\prime} \sin (\Delta m t) ~.
\end{equation}
The quantities $a_{\epsilon^\prime}$ and $a_{\epsilon + 
\epsilon^\prime}$ depend on 
the hadronic matrix elements:
\begin{equation}\label{a1}
a_{\epsilon^\prime} = \frac{ 1- |\lambda_{CP}|^2}{1+ |\lambda_{CP}|^2},
~~~a_{\epsilon+\epsilon^\prime }= \frac{ - 2Im(\lambda_{CP})}
{1+ |\lambda_{CP}|^2} ,
\end{equation}
where
\begin{equation}
\lambda_{CP} = \frac{V_{tb}^* V_{td}}{V_{tb} V_{td}^*}
\frac{\langle f|H_{eff}|\bar B^0\rangle  }{\langle f|H_{eff}| B^0\rangle  }. 
\end{equation}

For case (i) decays, the coefficient $a_{\epsilon^\prime}$ determines
$A_{CP}(t)$, and since no mixing is involved for these decays, the
CP-violating asymmetry is independent of time. We shall call these,
together with the
CP asymmetries in charged $B^\pm$ decays, CP-class (i) decays. 
For case (ii) and (iii), 
one has to separate the $\sin(\Delta m t)$ and  $\cos (\Delta m t)$
terms to get the CP-violating asymmetry $A_{CP}(t)$.
The time-integrated asymmetries are:
\begin{equation}
A_{CP} =\frac{1}{1+x^2} a_{\epsilon^\prime }
+\frac{x}{1+x^2} a_{\epsilon+\epsilon^\prime },\label{cpint}
\end{equation}
with $x=\Delta m /\Gamma \simeq 0.73$ for the $B^0$ - $\overline{B^0}$
case. 

Case (iv) also involves mixing but here one has to study the four 
time-dependent decay widths for $B^0(t) \to f$, $\bar B^0(t) \to \bar f$, 
$B^0(t) \to \bar f$ and $\bar B^0(t) \to  f$ \cite{class-four}. 
These time-dependent widths can be expressed by four basic matrix elements
\begin{equation}
\begin{array}{ll}
g=\langle f|H_{eff} |B^0\rangle  ,& h=\langle f|H_{eff}|\bar B^0\rangle ,\\
\bar g=\langle \bar f|H_{eff} |\bar B^0\rangle  ,
& \bar h=\langle \bar f|H_{eff}| B^0\rangle ,
\end{array}
\end{equation}
which determine the decay matrix elements of $B^0\to f \& \bar f$ and 
of $\bar B^0 \to \bar f \& f$ at $t=0$.
By measuring the time-dependent spectrum of the decay rates of $B^0$ and 
$\bar B^0$, one can find the coefficients of the two functions $\cos\Delta mt$
and $\sin \Delta m t$ and extract the quantities $a_{\epsilon '}$,
$a_{\epsilon +\epsilon '}$, $|g|^2+|h|^2$, $a_{\bar \epsilon '}$, 
$a_{\epsilon +\bar \epsilon '}$ and $|\bar g|^2+|\bar h|^2$ as well as 
$\Delta m$ and $\Gamma$.

  Estimates of $A_{CP}(\optbar{B^0} \to h_1 h_2)$,
representing the decays belonging to the CP-classes (i) to (iv), together 
with the branching 
ratios averaged over the charge-conjugated modes, are given in Table 2.
They have estimated branching ratios in excess of $10^{-6}$ (except for
the decay $\optbar{B^0} \to \pi^0 \pi^0$).
They also include the decay modes $\optbar{B^0} \to K^\pm \pi^\mp$,
$\optbar{B^0} \to \eta^\prime K_S^0 $, 
$\optbar{B^0} \to \pi^\pm K^{*\mp}$, 
$B^0/\bar{B}^0 \to \rho^-\pi^+/\rho^+\pi^-$,
$B^0/\bar{B}^0 \to \rho^+\pi^-/\rho^-\pi^+$, whose branching ratios have 
been measured by the CLEO collaboration. The CP asymmetries in all these
partial decay rates are expected to be large.   

\begin{table}[t]
\caption{Branching ratios measured by the CLEO collaboration
\protect\cite{Gao-Wuerthwein-99} and
factorization-based theoretical estimates of the same 
\protect\cite{AKL98-1}
(in units of $10^{-5}$), updated for the central values
of the CKM fits \protect\cite{AL99-1} $\rho=0.20$, $\eta=0.37$
and the factorization model parameter $\xi=0.5$. Theory numbers
correspond to
using the BSW model \protect\cite{BSW85} [Lattice QCD/QCD sum rule] for 
the
form factors. \label{tab:exp}}
\begin{center}
\footnotesize
\begin{tabular}{|c|c|c|l|}
\hline
Decay Mode &\raisebox{0pt}[13pt][7pt]{BR
(Expt)\protect\cite{Gao-Wuerthwein-99}} &
\raisebox{0pt}[13pt][7pt]{BR(Theory)\protect\cite{AKL98-1}}\\ \hline
$B^0 \to K^+ \pi^-$   &$1.4\pm 0.3 \pm 0.2$ & $1.4[1.7]$\\
\hline
$B^+ \to K^+ \pi^0$     &$1.5\pm 0.4 \pm 0.3$ & $1.0 [1.1] $\\
\hline
$B^+ \to K^0\pi^+$      &$1.4 \pm 0.5 \pm 0.2$ & $1.6 [1.9] $\\
\hline
$B^+ \to  \eta^\prime K^+$ &$7.4^{+0.8}_{-1.3} \pm 1.0$ &$ 2.3 
[2.7] $\\
\hline
$B^0 \to \eta^\prime K^0 $ &$5.9^{+1.8}_{-1.6}\pm 0.9$ &$ 2.3 [2.7] $\\
\hline
$B^0 \to \pi^- K^{*+} $ &$2.2 ^{+0.8 ~~+0.4}_{-0.6~~-0.5}$ & $ 0.6
[0.7] $\\
\hline 
$B^+ \to \pi^+ \rho^0$ &$1.5 \pm 0.5 \pm 0.4 $ &$ 0.9 [1.0] $\\
\hline
$B^0/\bar{B}^0 \to \rho^- \pi^+$ &$3.5^{+1.1}_{-1.0} \pm 0.5$ &$3.7 
[4.3]$\\ \hline
\end{tabular}
\end{center}
\end{table}

\section{Comparison of the Factorization Model with the CLEO 
Data} %
Before we discuss the numerical results,
a technical remark on the underlying theoretical framework  
is in order.
The estimates being discussed here \cite{AKL98-1,AKL98-2}, and the earlier 
work along these
lines \cite{AG97,ACGK97}, are all based on using the NLO virtual corrections
for the matrix elements of the four-quark operators, calculated in the 
Landau gauge with off-shell quarks \cite{Buras-Marti-92}. This renders the
effective (phenomenological) coefficients used in these works 
both gauge-and quark (off-shell) mass-dependent \cite{BS98-2}. The remedy 
for this unsatisfactory situation is to replace the
virtual corrections with the ones calculated with on-shell quarks,
which are manifestly gauge invariant. A proof that gauge- and 
renormalization-scheme-independent effective coefficients (and hence
decay amplitudes) can be consistently obtained in perturbation theory 
now exists \cite{Cheng99-1}. Further discussion
of these aspects and alternative derivation of the gauge-invariant on-shell
amplitudes in $B \to h_1 h_2$ decays will be reported elsewhere
\cite{ACGK99-1}. From the phenomenological point of view, the
corrections in the effective coefficients are, however, 
small \cite{Chenetal99-1,AKL99-1}. So, the real big unknown in this 
approach is the influence of the neglected soft non-factorizing 
contributions.
 
 We now compare the predictions of the
factorization-based estimates with the CLEO data.
Since no CP asymmetries in the decays $B \to h_1h_2$ have so far been
measured, this comparison can only be done in terms of the branching ratios.
We give in Table 3  eight $B \to h_1h_2$ decay modes, their branching 
ratios measured by CLEO \cite{Gao-Wuerthwein-99} and the updated 
theoretical estimates of the same \cite{AKL98-1}. 
 The entries
given for $B^0/\bar{B}^0 \to \rho^- \pi^+$ are for the sum of the two modes,
as CLEO does not distinguish between the decay products of $B^0$ and 
$\bar{B}^0$. Moreover, we have fixed
the factorization model parameter to $\xi=0.5$, which is suggested
by the measured $B \to \pi \rho$ branching ratios. Table 3
shows that all five $K\pi$ and $\pi\rho$ decay modes are well-accounted for
in the factorization-model,
but the $\pi^\pm K^{*\mp}$  and the two $B \to \eta^\prime K$ modes are 
typically a
factor 2 below the CLEO data, though experimental errors are large to 
be conclusive. Also, there is some dependence of the decay rates on the form 
factors,
as indicated by the two set of numbers corresponding to the use of the
BSW-model \cite{BSW85} and the Lattice-QCD/QCD-sum-rule-based 
estimates \cite{AKL98-1}, and on $\xi$.

 While the final verdict of the
experimental jury is not yet in, it is fair to say 
that the factorization approach, embedded 
in a well-defined perturbative framework, provides a description of the
present data which is accurate within a factor 2. This is a hint
that non-factorizing soft final state interactions do not represent 
a dominant theme in $B$ decays. Present data gives some 
credibility to the idea that perturbative QCD-based methods can be
used in estimating final state interactions.
Non-leptonic $B \to h_1 h_2$ decays discussed here, many of which will
be measured in experiments at the B factories and hadron machines,  will 
test this quantitatively.

\noindent
{\bf Acknowledgements}:

 I thank Gustav Kramer, David London and Cai-Dian L\"u for very  
productive collaborations, many helpful discussions and input to this report.
Thanks are also due to Frank W\"urthwein for the communication and 
discussion of 
the CLEO data. The warm hospitality of the organizers of the Hadron13
conference  is thankfully acknowledged. 


\end{document}